\documentclass[useAMS,usenatbib]{mn2e}
\usepackage{epsfig,color}
\def\reff@jnl#1{{\rm#1\/}}

\def\aj{\reff@jnl{AJ}}                  
\def\araa{\reff@jnl{ARA\&A}}            
\def\apj{\reff@jnl{ApJ}}                
\def\apjl{\reff@jnl{ApJ}}               
\def\apjs{\reff@jnl{ApJS}}              
\def\apss{\reff@jnl{Ap\&SS}}            
\def\aap{\reff@jnl{A\&A}}               
\def\aapr{\reff@jnl{A\&A~Rev.}}         
\def\aaps{\reff@jnl{A\&AS}}             
\def\baas{\reff@jnl{BAAS}}              
\def\jrasc{\reff@jnl{JRASC}}            
\def\memras{\reff@jnl{MmRAS}}           
\def\mnras{\reff@jnl{MNRAS}}            
\def\physrep{\reff@jnl{Phys.Rep.}}
\def\pra{\reff@jnl{Phys.Rev.A}}         
\def\prb{\reff@jnl{Phys.Rev.B}}         
\def\prc{\reff@jnl{Phys.Rev.C}}         
\def\prd{\reff@jnl{Phys.Rev.D}}         
\def\prl{\reff@jnl{Phys.Rev.Lett}}      
\def\pasp{\reff@jnl{PASP}}              
\def\pasj{\reff@jnl{PASJ}}              
\def\skytel{\reff@jnl{S\&T}}            
\def\solphys{\reff@jnl{Solar~Phys.}}    
\def\sovast{\reff@jnl{Soviet~Ast.}}     
\def\ssr{\reff@jnl{Space~Sci.Rev.}}     
\def\nat{\reff@jnl{Nature}}             

\title[Clustering with H{\sc i}]{The clustering of galaxies as a function of
  their photometrically-estimated atomic gas content}

\author[C. Li et al.]{Cheng Li$^{1}$\thanks{E-mail: leech@shao.ac.cn},
  Guinevere Kauffmann$^{2}$, Jian Fu$^{2,1}$, Jing Wang$^{2,3}$,
  Barbara Catinella$^{2}$,  \newauthor Silvia Fabello$^{2}$,
  David Schiminovich$^{4}$ and Wei Zhang$^{5}$\\
  $^{1}$Partner Group of the Max Planck Institute for Astrophysics
  and Key Laboratory for Research in Galaxies and Cosmology \\
  ~of Chinese Academy  of  Sciences, Shanghai Astronomical Observatory,
  Nandan   Road  80,  Shanghai  200030,  China \\
  $^{2}$Max-Planck-Institute f\"{u}r Astrophysik,
  Karl-Schwarzschild-Str.  1, D-85741 Garching, Germany \\
  $^{3}$Department of Astronomy, University of Science and
  Technology of China, Jinzhai Road 96, Hefei 230026, China \\
  $^{4}$Department of Astronomy, Columbia University,
  550 West 120th Street, New York, NY 10027, USA \\
  $^{5}$National Astronomical Observatories, Chinese Academy 
  of Sciences, Beijing 100012, China}

\begin{document}

\date{Accepted ........ Received ........; in original form ........}

\pagerange{\pageref{firstpage}--\pageref{lastpage}} \pubyear{2012}

\maketitle

\label{firstpage}

\begin{abstract}
We introduce a new photometric estimator of the H{\sc i} mass fraction
($M_{HI}/M_\ast$) in local galaxies,  which is a linear combination of
four parameters: stellar mass,  stellar surface mass density , $NUV-r$
colour, and $g-i$  colour gradient. It is calibrated  using samples of
nearby galaxies  ($0.025<z<0.05$) with  H{\sc i} line  detections from
the  GASS and  ALFALFA  surveys,  and it  is  demonstrated to  provide
unbiased  $M_{HI}/M_\ast$ estimates even  for H{\sc  i}-rich galaxies.
We apply this estimator to  a sample of $\sim$24,000 galaxies from the
SDSS/DR7 in  the same  redshift range. We  then bin these  galaxies by
stellar  mass and  H{\sc i}  mass fraction  and compute  projected two
point cross-correlation  functions with respect to  a reference galaxy
sample.    Results  are   compared  with   predictions   from  current
semi-analytic models  of galaxy formation.  The agreement is  good for
galaxies with stellar masses  larger than $10^{10} M_{\odot}$, but not
for lower mass systems.

We then extend  the analysis by studying the bias in the clustering of 
H{\sc i}-poor or H{\sc i}-rich galaxies  with respect to galaxies with
normal H{\sc i} content on  scales between 100  kpc and $\sim  5$ Mpc. 
 For  the H{\sc i}-deficient population, the strongest  bias effects arise 
when the H{\sc i}-deficiency  is defined  in  comparison to  galaxies  of the  same
stellar mass  and size.  This  is not reproduced by  the semi-analytic
models, where the quenching of  star formation in satellites occurs by
``starvation'' and does not depend on their internal structure.  H{\sc
  i}-rich galaxies  with masses  greater than $10^{10}  M_{\odot}$ are
found to be anti-biased compared  to galaxies with ``normal'' H{\sc i}
content.  Interestingly,  no  such  effect  is found  for  lower  mass
galaxies.
\end{abstract}


\section{Introduction}\label{S:introduction}

Over the past decade, large  optical spectroscopic surveys such as the
2dF  Galaxy  Redshift Survey  \citep[2dFGRS;][]  {Colless-01} and  the
Sloan  Digital  Sky  Survey  \citep[SDSS;][]{York-00} have  led  to  a
resurgence in  studies of  the large scale  structure of  the Universe
traced by galaxies.  There are  two main applications of such studies:
a) to constrain cosmological parameters  such as the matter density of
the Universe  $\Omega_m$, Hubble parameter  $h$, fluctuation amplitude
$\sigma_8$  and neutrino  mass  in conjunction  with constraints  from
other experiments, such as  cosmic microwave background (CMB) or Lyman
$\alpha$  forest measurements  \citep[e.g.][]{Spergel-03, Tegmark-04b,
  Eisenstein-05},  b)  to  constrain  models  for  the  formation  and
evolution of the galaxy population.

Traditionally,   large-scale   structure   studies   that   focus   on
cosmological  applications aim  to  measure the  clustering signal  on
large  scales  (tens  of  Mpc  or  greater).   On  large  scales,  the
clustering amplitude depends only on the mass of the dark matter halos
that  host the  galaxies. All  galaxies, regardless  of mass  or type,
trace  the underlying  dark matter  density field  in a  simple linear
fashion, so constraints on  cosmological parameters are believed to be
robust.

In contrast,  studies aimed at constraining galaxy  formation focus on
the  clustering signal on  scales less  than $\sim  5$ Mpc.   On these
scales, the clustering  amplitude depends on not only  the mass of the
dark  matter halos  in which  galaxies are  found, but  also  the {\em
  location}     of     galaxies     within    their     host     halos
\citep[]{Benson-00b, Peacock-Smith-00}.

In the current paradigm of galaxy formation within a merging hierarchy
of dark matter halos, galaxies form when gas is able to cool, condense
and form stars at the {\em centres} of dark matter haloes.  At a later
stage, the galaxy  may be accreted into a larger  dark matter halo and
become a  {\em satellite} galaxy in a  group or a cluster.   Gas is no
longer  supplied to  these  galaxies and  star formation  subsequently
shuts down over  some timescale \citep{ Kauffmann-White-Guiderdoni-93,
  Cole-94}.  In more  recent  models,  gas is  no  longer supplied  to
central  galaxies with  central super-massive  black holes  located in
dark matter halos  containing a hot gas atmosphere  
\citep[]{Croton-06, Bower-06}.

One important  goal in  modern galaxy formation  is to  understand the
physics  behind   these  gas-related  ``accretion"   and  ``quenching"
processes in  detail, because the  timescales over which  they operate
and the way in which  their efficiencies scale with halo and/or galaxy
mass will determine how the galaxy  population as a whole evolves as a
function of  cosmic epoch. Clustering  analysis is a powerful  tool in
this endeavour. In particular,  analysis of the {\em cross-correlation}
between a  specific galaxy  sub-population and a  larger ``reference''
sample allows one to maximize  the $S/N$ of the clustering measurement
when the size of the  sub-sample is small.
This technique has recently been applied to sub-samples of narrow-line
galaxies  with   actively  accreting  black  holes  in   the  SDSS  to
demonstrate that these  galaxies are not triggered by  mergers and are
found  preferentially  at the  centres  of  their  dark matter  haloes
\citep[][]{Li-06c,Li-08a}.

The clustering of galaxies as  a function of their neutral gas content
should  in  principle  yield   very  interesting  constraints  on  gas
accretion      and      quenching      processes      in      galaxies
\citep[e.g.][]{Popping-09,  Kim-11}.  \citet{Meyer-07}  determined the
two-point  autocorrelation function (2PCF)  of H{\sc  i}-rich galaxies
using  4315  galaxies  from  the   H{\sc  i}  Parkes  All  Sky  Survey
\citep[HIPASS;][]{Zwaan-05} and found  that H{\sc i} selected galaxies
exhibit weaker clustering than optically selected galaxies of the same
luminosity.  Recently \citet{Passmoor-Cress-Faltenbacher-11a} measured
the  2PCF  for  an early  release  of  the  Arecibo Legacy  Fast  ALFA
\citep[ALFALFA;][]    {Giovanelli-05}    sample,    finding    similar
results. The  \citet{Meyer-07} study also looked at  the dependence of
clustering on  total H{\sc i} mass,  finding it to be  weaker than the
dependence on both luminosity and on rotation velocity.

Up to now,  there has been no attempt to  study how clustering depends
on H{\sc i} mass {\em fraction} (i.e $M_{HI}/M_\ast$), a quantity that
ought  to be  much more  directly related  to accretion  and quenching
processes that affect the gas content of a galaxy, but not its stellar
mass.  In addition, a power-law  form for the correlation function has
been {\em assumed} in  these previous clustering analyses, which means
that  information about  location of  the galaxies  within  their dark
matter  haloes (alternatively central  or satellite  galaxy fraction),
cannot be recovered.  Finally, because available samples are small, it
has not  been possible to study  clustering as a function  of H{\sc i}
mass  fraction in conjunction  with other  galaxy parameters,  such as
stellar mass  or stellar surface mass  density.  In this  work we will
demonstrate how an  approach that combines H{\sc i}  data for a small,
but  complete sample  of 1000  galaxies and  optical data  for  a much
larger   sample   of  galaxies   from   the   SDSS   Data  Release   7
\citep[DR7;][]{Abazajian-09}  can be  used to  study the  influence of
dark matter  halo mass  and environment on  the gaseous  properties of
galaxies.

The  GALEX   Arecibo  SDSS  Survey   \citep[GASS;][]{Catinella-10}  is
measuring the  atomic gas content  of a sample of  $\sim$1000 galaxies
with  redshifts and stellar  masses in  the ranges  $0.025<z<0.05$ and
$10^{10}<M_\ast<10^{11.5}M_\odot$.  Each galaxy  is observed until the
H{\sc i}  line is detected or  until an upper limit  of $\sim0.015$ in
the atomic-to-stellar  mass ratio is  reached.  The GASS  galaxies are
selected  from the  SDSS spectroscopic  and Galaxy  Evolution Explorer
\citep[GALEX;][]{Martin-05}, so  stellar masses, sizes  and structural
parameters    are    available   from    the    MPA/JHU   data    base
(http://www.mpa-garching.mpg.de/SDSS/).  The  scaling relations of the
H{\sc i}  mass fraction of  the GASS galaxies ($M_{HI}/M_\ast$),  as a
function of  global galaxy parameters  such as stellar  mass $M_\ast$,
surface  mass  density   $\mu_\ast$,  light  concentration  index  $C$
(defined as $R_{90}/R_{50}$,  the ratio of the radii  enclosing 90 and
50 percent  of the total  $r$-band light) and specific  star formation
rate   $SFR/M_\ast$   are   presented   in   \citet[][hereafter   C10]
{Catinella-10} and \citet[][]{Schiminovich-10}.

Following the  work of  \citet{Zhang-09b}, C10 defined  a gas-fraction
``plane'' linking H{\sc i} mass fraction, stellar surface mass density
and  $NUV-r$  colour  that  exhibited   a  scatter  of  0.315  dex  in
$\log_{10}M_{HI}/M_\ast$,  considerably   tighter  that  the  relation
between H{\sc  i} mass fraction and optical/near-IR  colour studied by
\citet{Kannappan-04},  which had  a scatter  of $\sim  0.4$  dex.  The
improvement in scatter indicates that the H{\sc i} content of a galaxy
scales with its physical size as well as with its star formation rate.
In  subsequent work,  \citet{Wang-11b}  showed that  at fixed  $NUV-r$
colour and stellar surface density,  galaxies with larger H{\sc i} gas
fractions have bluer outer disks.

In this paper, we include the  {\em colour gradient} of galaxies as an
additional  parameter in  our  fits.  This  produces  a relation  with
similar scatter, but  that better predicts the H{\sc  i} mass fraction
of the most gas-rich galaxies in our samples.  We use this relation to
predict the H{\sc  i} content of the galaxies  in our SDSS/DR7 sample.
We then study how clustering  depends on both ``pseudo'' H{\sc i} mass
fraction, and a ``pseudo'' H{\sc i} excess/deficiency parameter, which
we define  as the  deviation in  the predicted H{\sc  i} content  of a
galaxy from the  average H{\sc i} content of all  galaxies of the same
stellar  mass and  surface  mass density.   This  ``pseudo'' H{\sc  i}
excess/deficiency parameter depends on a combination of $NUV-r$ colour
and  $g-i$ colour  gradient.   Finally, we  compare  our results  with
clustering predictions from  the semi-analytic galaxy formation models
of     \citet[][hereafter    F10]{Fu-10}     and    \citet[][hereafter
  G11]{Guo-11b}.  The main way in which the F10 model differs from the
G11 model is  that it includes simple prescriptions  for molecular gas
formation processes.

The motivation behind expressing the results in this paper in terms of
``pseudo''  H{\sc  i}  fraction,  rather  than in  terms  of  directly
measured photometric quantities, is because this provides insight into
physical  processes  regulating  the  gas  supply  in  galaxies.   The
semi-analytic  models make  a host  of  assumptions about  how gas  is
accreted from  the surround  dark matter halo  and then  consumed into
stars. By  comparing clustering as a  function of gas  fraction in the
models with the  data, we hope to ascertain  whether these assumptions
are correct,  or whether there are discrepancies  that warrant further
investigation.

Throughout this  paper we have assumed a  cosmology with $\Omega=0.3$,
$\Lambda=0.7$   and  $H_0=70$   kms$^{-1}$Mpc$^{-1}$   when  computing
observed    quantities.     A    Hubble    constant    of    $H_0=100$
kms$^{-1}$Mpc$^{-1}$ is  assumed when presenting  correlation function
measurements.   We note  that  the F10  and  G11 models  are based  on
simulations with  $\Omega=0.25$ and $\Lambda=0.75$.  This  will make a
small difference  in the comparison  between data and models.  We note
that the focus of this paper is not on obtaining precision fits to the
data,  but on  identifying major  discrepancies  that may  lead us  to
change the input physics in the model.

\section{Data} \label{sec:data}

\subsection{GASS galaxy sample} \label{sec:gasssample}

The parent  sample of GASS  consists of 12,006 galaxies  selected from
the    region    of    sky    where    the    sixth    data    release
\citep[DR6;][]{Adelman-McCarthy-08} of  the SDSS overlaps  the maximal
ALFALFA footprint.   All galaxies are selected to  have stellar masses
$M_\ast>10^{10}M_\odot$  and redshifts  in  the range  $0.025<z<0.05$.
The  GASS sample  is constructed  by  randomly selecting  a subset  of
$\sim$1000 galaxies from the parent sample within the footprint of the
GALEX Medium Imaging  Survey so that the distribution  in stellar mass
is flat.   The targets are  observed with the Arecibo  radio telescope
until  detected or  until an  H{\sc i}  mass  fraction $M_{HI}/M_\ast$
limit of  1.5-5 per cent  is reached.  In  this work, we use  the {\em
  representative}   sample  of  480   GASS  galaxies,   including  293
detections and  187 non-detections described  in \citet{Catinella-11}.
Details  of the  GASS survey  design, target  selection  and observing
procedures can be found in C10.

\subsection{SDSS galaxy samples} \label{sec:sdsssample}

We have constructed two galaxy samples from the SDSS/DR7.

The first  sample, which will serve  as our {\em  reference} sample in
the  clustering  analysis, is  a  magnitude-limited  sample of  66,461
galaxies  with $r<17.6$,  $-24<M_{^{0.1}r}<-16$ and  redshifts  in the
range $0.025<z<0.05$.   Here, $r$  is the $r$-band  Petrosian apparent
magnitude, corrected for Galactic extinction, and $M_{^{0.1}r}$ is the
$r$-band  Petrosian absolute  magnitude, corrected  for  evolution and
$K$-corrected to its value  at $z=0.1$. These selection criteria, with
the exception of  the redshift range, are the same  as in our previous
papers  where  we studied  the  clustering  of  galaxy luminosity  and
stellar  mass \citep[e.g.][]{Li-12}.  We  have generated  {\em random}
samples that have the same sky  coverage as well as the same position-
and  redshift-dependent  selection effects  as  the reference  sample.
Details  of  this  procedure  are  presented in  our  previous  papers
\citep[e.g.][]{Li-06a}.

The second  sample contains  36,136 galaxies, and  is a subset  of the
reference    galaxies    with   stellar    masses    in   the    range
$10^{9.5}M_\odot<M_\ast<10^{11}M_\odot$.  In the  next section we will
estimate  an H{\sc i}  mass fraction  for each  galaxy in  this sample
using our newly calibrated  photometric estimator.  We select a number
of subsamples  binned according  to $M_\ast$ and  $M_{HI}/M_\ast$, and
cross-correlate these with both the reference and the random samples.

\subsection{Physical properties of galaxies} \label{sec:quantities}

The physical  quantities necessary for this work  include stellar mass
$M_\ast$, stellar surface mass  density $\mu_\ast$, $NUV-r$ colour and
colour gradient $\Delta_{g-i}$.  Stellar  masses are derived from SDSS
photometry using the methodology described in \citet{Salim-07}.  These
masses       are        publically       available       at       {\it
  http://www.mpa-garching.mpg.de/SDSS/DR7}.  The  stellar surface mass
density  is  defined  as  $\mu_\ast=M_\ast/(2\pi  R^2_{50,z})$,  where
$R_{50,z}$ is the physical radius in units of kpc that contains half the
total  light in  the $z$-band. The $NUV$ magnitude is provided by the
GALEX pipeline and the  $NUV-r$ colour  is corrected  for
Galactic      extinction      following     \citet{Wyder-07}      with
$A_{NUV-r}=1.9807A_r$,  where  $A_r$  is  the extinction  in  $r$-band
derived  from the  dust maps  of \citet{Schlegel-Finkbeiner-Davis-98}.
The $g-i$ colour gradient is  defined as $\Delta_{g-i} = (g-i)_{out} -
(g-i)_{in}$,  where  $(g-i)_{in}$  and  $(g-i)_{out}$  are  the  $g-i$
colours  in the  inner and  outer regions  of the  galaxy.   The inner
region  is enclosed  by  $R_{50,r}$, the  radius  containing half  the
$r$-band  light.  The  outer region  is  defined as  the area  between
$R_{50}$  and  $R_{90}$, the  radius  enclosing  90  per cent  of  the
$r$-band light.   A negative value of $\Delta_{g-i}$  implies that the
outer region of the galaxy is {\em bluer} than the inner region.

\subsection{Semi-analytic model galaxy catalogues and mock SDSS samples}
\label{sec:samcatalogues}

In this paper we compare our observational results to predictions from
the galaxy formation models of  G11 and F10.  Both models were created
by  implementing  simple prescriptions  for  baryonic astrophysics  on
merger trees that follow  the evolution of the halo/subhalo population
in  the  Millennium  \citep[MS;][]{Springel-05b} Simulation,  a  cubic
region 500 $h^{-1}$Mpc  on a side with mass  resolution $\sim 10^{10}$
M$_\odot$.

The G11 model  is the most recent semi-analytic  model from the Munich
group,  in which the  treatments of  supernova feedback,  galaxy size,
photoionization  suppression and  environmental  effects on  satellite
galaxies have been significantly  updated. G11 demonstrated that their
model  provided excellent  fits not  only  to the  the luminosity  and
stellar mass functions of galaxies derived from SDSS data, but also to
recent  determinations of  the abundance  of faint  satellite galaxies
around  the Milky  Way.  The  clustering properties  of galaxies  as a
function of stellar mass predicted  by the model are in good agreement
with  SDSS  data for  masses  above  $6\times  10^{10}M_\odot$ and  at
separations larger than 2 Mpc.  On smaller scales, lower mass galaxies
are predicted to be substantially more clustered than observed.

The  F10  model  is  based   on  an  earlier  version  of  the  Munich
semi-analytic code, which is  described in detail in \citet{Croton-06}
and updated in \citet[][hereafter DB07]{DeLucia-Blaizot-07}.  The main
new aspect of  this model is that galactic discs  are represented by a
series of concentric rings in order  to track the evolution in the gas
and stellar surface density profiles of galaxies over cosmic time.  In
addition,  two  simple   prescriptions  for  molecular  gas  formation
processes are included:  one is based on the  analytic calculations by
\citet{Krumholz-McKee-Tumlinson-09a}, and one  is a prescription where
the H$_2$ fraction  is determined by the pressure  of the interstellar
medium  \citep{Blitz-Rosolowsky-06}.   The  model is  currently  being
configured to  operate on the latest  code of G11.   The comparison in
this paper will be restricted to  the model published in the F10 paper
and      to     the      H$_2$      formation     prescription      of
\citet{Krumholz-McKee-Tumlinson-09a}.

In this  paper, the different treatments  of gas stripping  in the G11
and F10  models are of interest  to us.  In most  SAMs including DB07,
hot gas in a halo is assumed to be stripped immediately after the halo
has  been  accreted  on to  a  larger  halo.  In  the G11  model  this
prescription  has been  modified.  Satellite  galaxies that  still are
attached to  a sub-halo within  the larger virialized  ``parent'' halo
are still  able to accrete gas.   This new treatment  was motivated by
observational findings  and hydrodynamical simulations  which revealed
that the hot atmosphere of  massive satellite galaxies may survive for
a considerable  time after accretion  (see G11 and  references therein
for details). This change primarily affects satellite galaxies located
in the  outer regions of their  host dark matter  halos. The timescale
for  gas   to  be  depleted   and  star  formation  to   stop  becomes
significantly longer.

We have constructed  a set of 50 mock SDSS  galaxy catalogues from the
G11  model using  both the  sky mask  and the  magnitude  and redshift
limits  of our  SDSS  reference sample.  Detailed  description of  our
methodology can  be found in \citet{Li-06c}  and \citet{Li-07b}. These
mock catalogues allow  us to derive realistic error  estimates for the
statistics measured below, including both sampling and cosmic variance
uncertainties.

\section{Estimating H{\sc i} mass fractions for the SDSS galaxies}
\label{sec:hiestimator}

\begin{figure}
\begin{center}
\epsfig{figure=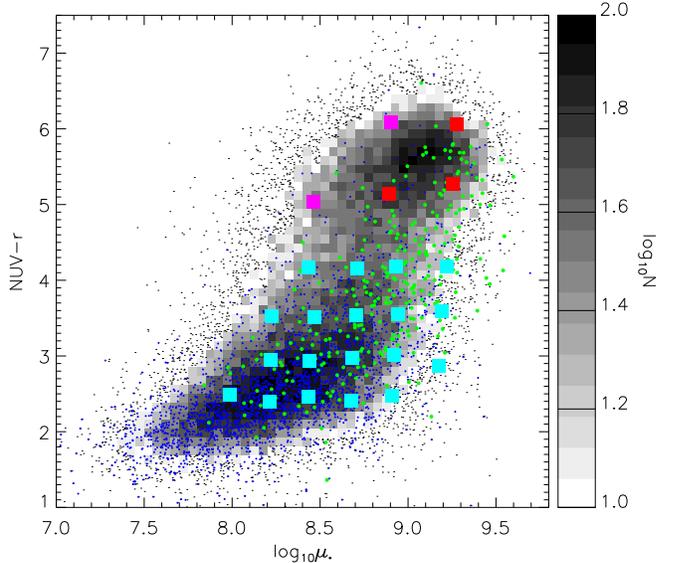,clip=true,width=0.5\textwidth}
\end{center}
\caption{Our different galaxy samples  in the two dimensional plane of
  NUV-r  colour and  stellar  surface mass  density.   The grey  scale
  indicates  the   location  of  the  36,136  galaxies   with  $9.5  <
  \log_{10}(M_\ast/M_\odot) < 11$ from  the SDSS DR7. The green points
  are  GASS galaxies  with H{\sc  i} detections.  The blue  points are
  ALFALFA-detected  galaxies in the  same redshift  range as  the GASS
  sample ($0.025 <  z < 0.05$). The squares  indicate the grid centers
  of  the  H{\sc  i}   stacking  analysis  of  an  optically-selected,
  volume-limited   sample   of    5000   galaxies   carried   out   by
  \citet{Fabello-11}.  The cyan and  red squares indicate the areas of
  the grid where a high  significance measurement of the mean H{\sc i}
  mass  fraction could  be obtained  from the  stacked  spectrum.  The
  purple squares indicate the regions  where only an upper limit could
  be derived.}
  \label{fig:mustarnuvr}
\end{figure}

\begin{figure*}
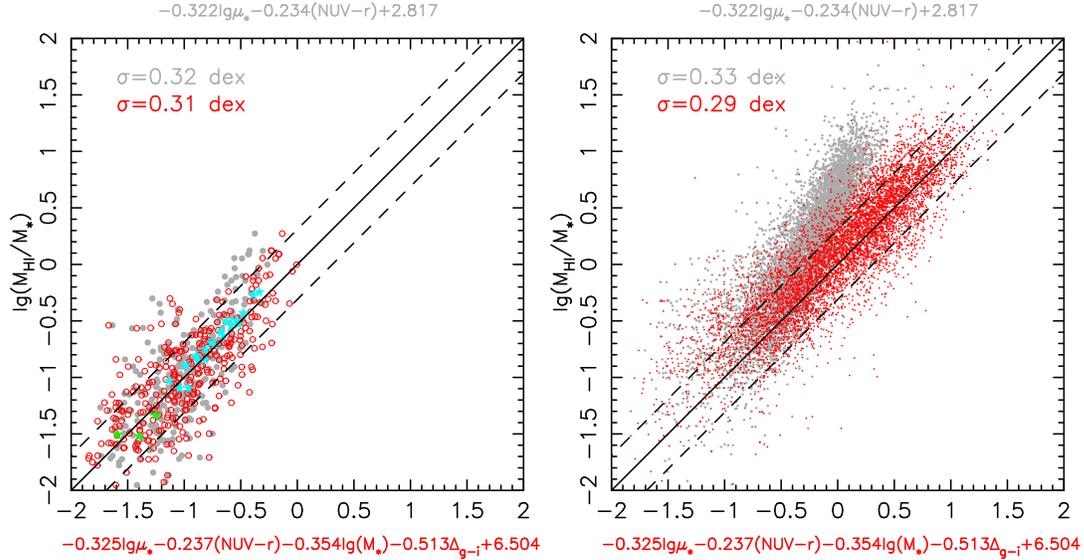

  \begin{center}
    \epsfig{figure=f2a.ps,clip=true,width=0.4\textwidth}
    \epsfig{figure=f2b.ps,clip=true,width=0.4\textwidth}
  \end{center}
  \caption{In  the left-hand  panel we  plot in  red open  circles the
    best-fitting   relation  between  the   H{\sc  i}   mass  fraction
    ($M_{HI}/M_\ast$)  and  the  linear  combination of  surface  mass
    density $\mu_\ast$, $NUV-r$ colour,  stellar mass $M_\ast$ and the
    gradient  in  $g-i$ colour  ($\Delta_{g-i}$)  determined from  the
    H{\sc i}-detected  galaxies in the  GASS survey.  The  relation is
    given in red on the bottom  of the panel.  This is compared to the
    grey dots which show the relation obtained by \citet{Catinella-10}
    between $M_{HI}/M_\ast$  and the linear  combination of $\mu_\ast$
    and $NUV-r$  (given in grey  on the top  of the panel).   Cyan and
    green diamonds are results for the stacked spectra (green diamonds
    are for the  3 stacks on the red  sequence).  The right-hand panel
    shows the same  thing for a sample of  H{\sc i}-rich galaxies from
    the ALFALFA  survey.  The  solid and dashed  lines in  both panels
    indicate the $1:1$ relation and 1$\sigma$ error region for the new
    estimator.}
  \label{fig:hiplane}
\end{figure*}

\begin{figure*}
  \begin{center}
    \epsfig{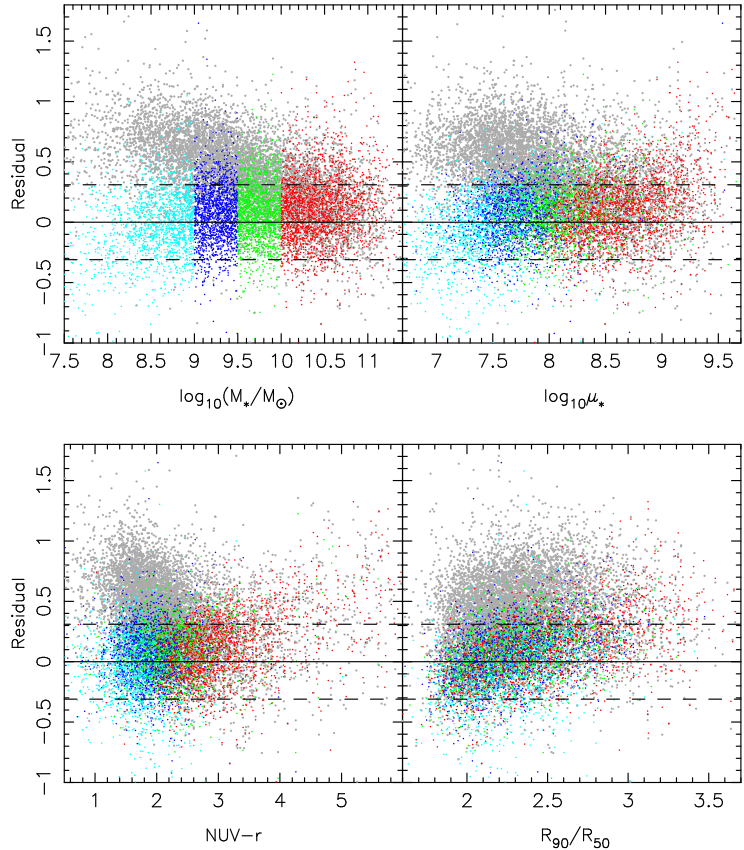}
  \end{center}
  \caption{Distributions   of   the   residuals   in   the   predicted
    $M_{HI}/M_\ast$ are plotted as functions of stellar mass $M_\ast$,
    stellar  surface  mass  density  $\mu_\ast$,  $NUV-r$  colour  and
    concentration   index  $R_{90}/R_{50}$   for   the  estimator   of
    \citet{Catinella-10}  (grey dots)  and  for the  new estimator  in
    equation  (\ref{eqn:hiplane}) (colourful dots). In  the latter
    case the galaxies are divided into four stellar mass intervals and
    are       plotted      in      different       colours      (cyan:
    $\log_{10}(M_\ast/M_\odot)<9$;                                blue:
    $9<\log_{10}(M_\ast/M_\odot)<9.5$;                           green:
    $9.5<\log_{10}(M_\ast/M_\odot)<10$;                            red:
    $\log_{10}(M_\ast/M_\odot)>10$). }
  \label{fig:residuals}
\end{figure*}

\begin{figure*}
  \begin{center}
    \epsfig{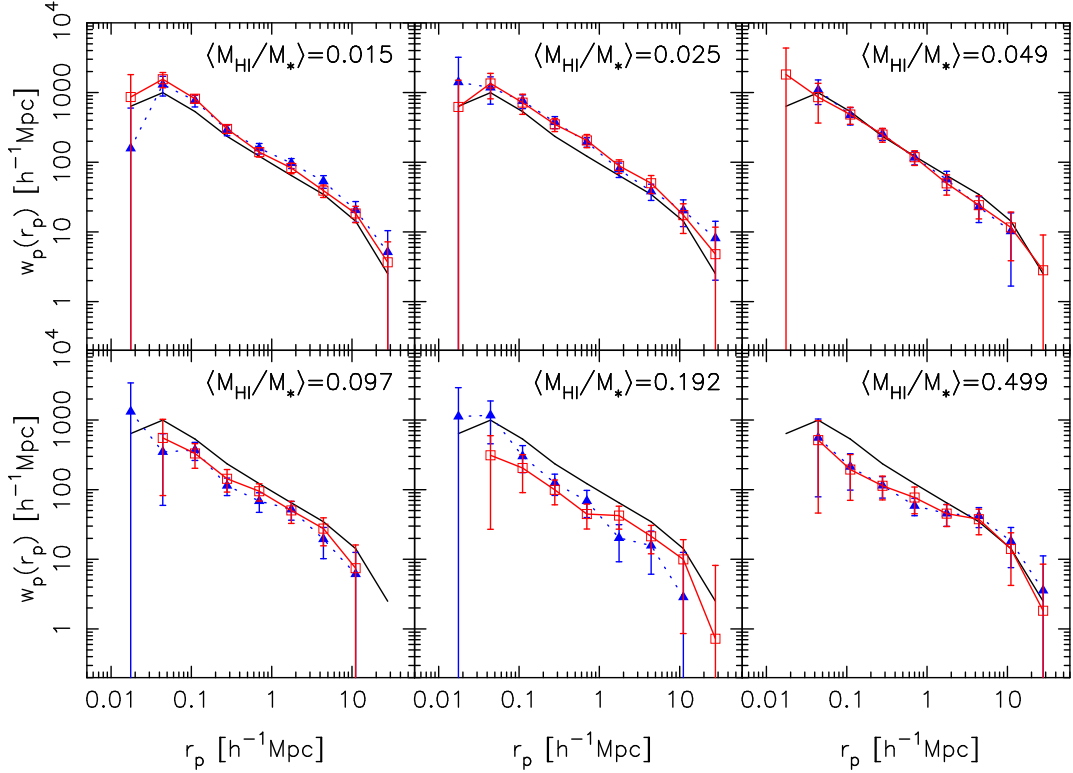}
  \end{center}
  \caption{Projected two-point cross-correlation function, $w_p(r_p)$,
    for six  subsamples with equal  number of galaxies from  the GASS,
    selected  by  the  observed  H{\sc  i}  mass  fraction  (triangles
    connected by  blue dotted lines)  or the predicted  value (squares
    connected by red solid lines).   The average value of the H{\sc i}
    mass fraction of  each subsample is indicated. The  result for the
    whole sample  is plotted in the  black solid line  and repeated in
    every panel for reference.}
  \label{fig:gasswrp}
\end{figure*}

There   have  been  a   number  of   attempts  to   calibrate  colours
\citep[e.g.][]{Kannappan-04}   or   emission-line  equivalent   widths
\citep[e.g.][]{Tremonti-04,  Erb-06b, Bouche-07}  as  proxies for  the
gas-to-stellar mass  ratio in galaxies.   \citet{Zhang-09b} proposed a
method   motivated  by  the   Kennicutt-Schmidt  star   formation  law
\citep{Schmidt-63,  Kennicutt-98a} that  combines  colour and  surface
brightness to estimate the H{\sc i}-to-stellar mass ratio. They used a
sample  of 800  galaxies  with  H{\sc i}  mass  measurements from  the
HyperLeda catalogue \citep{Paturel-03} and optical photometry from the
SDSS to  calibrate a relation linking these  quantities. In subsequent
work,  C10  used  an  unbiased   sample  of  galaxies  with  H{\sc  i}
measurements  from  GASS to  show  that  $M_{HI}/M_\ast$  can be  well
approximated  by   a  linear  combination   of  NUV-to-optical  colour
($NUV-r$)  and  stellar  surface  mass  density  ($\mu_\ast$)  with  a
$1\sigma$ scatter  of $\sim0.3$  dex. However, as  could be  seen from
Figure  12 of  C10, galaxies  detected by  the much  shallower ALFALFA
survey in redshift  range as the GASS sample  had significantly higher
H{\sc i}  mass fractions, and were also  systematically displaced from
the C10  plane.  This  result would  seem to imply  that the  H{\sc i}
masses of the  most gas-rich galaxies in the  local Universe cannot be
reliably inferred from their UV/optical properties.
 
However,  a recent  study by  \citet{Wang-11b} focusing  on  the H{\sc
  i}-rich galaxies from the GASS and ALFALFA samples has revealed that
unusually H{\sc i}-rich  galaxies have bluer-than-average outer disks.
Motivated  by this  finding,  we now  propose  an updated  photometric
estimator, that includes  both stellar mass and the  gradient in $g-i$
colour as additional parameters.  

In Figure~\ref{fig:mustarnuvr}, we plot  some of the galaxy samples we
will be  working with in  this paper in  the two dimensional  plane of
$NUV-r$  colour and  stellar  surface mass  density.   
There are 3 grid centers (plotted in red squares) located well
within the ``red sequence'' with  good mean H{\sc i} mass measurements
and  these provide  a  check on  whether  our H{\sc  i} mass  fraction
estimators work well in regime where galaxies are gas-poor on average.
As can be seen, the combination  of the GASS and ALFALFA data, as well
as  the   stacked  results,  cover   the  region  of   $NUV-r$  versus
$\log_{10}\mu_\ast$ parameter space reasonably well. The GASS galaxies
and stacked  results are offset  to somewhat higher values  of stellar
surface mass density, because these samples are restricted to galaxies
with stellar masses larger than $10^{10} M_{\odot}$.

Our new estimator is defined by
\begin{eqnarray}\label{eqn:hiplane}
  \log_{10}M_{HI}/M_\ast  &  =  &  a  \log_{10}\mu_\ast  +  b  (NUV-r)
  \nonumber \\ & & + c \log_{10}M_\ast/M_\odot + d \Delta_{g-i} + e
\end{eqnarray}
where   $\Delta_{g-i}$    is   the   colour    gradient   defined   in
\S~\ref{sec:quantities}. The coefficients are determined by minimizing
the residuals from the plane using  the 293 H{\sc i} detections in the
GASS sample.  Following C10, when carrying out the fit, we weight each
galaxy by  the mass-dependent selection  function of the  GASS survey.
The $1\sigma$ scatter  in our new estimator is  0.31 dex, very similar
to that of the old one.  Figure~\ref{fig:hiplane} illustrates how this
new estimator improves the H{\sc i} mass fraction estimates.

In the  left-hand panel  of the  figure, grey dots  show the  H{\sc i}
plane of C10 for GASS  galaxies, while coloured stars show the stacked
results  \footnote  {Note the  bins  on  the  red sequence  have  been
  coloured  in green and  lie very  close to  an extrapolation  of the
  best-fit line through the other  bins, indicating that the C10 plane
  still yields an  accurate prediction of mean H{\sc  i} mass fraction
  for galaxies  on the red sequence}.   In the right-hand  panel of the
figure, grey dots  show the same C10 plane for  a sample of $\sim7000$
galaxies  from  the  $\alpha.40$   catalogue  of  the  ALFALFA  survey
\citep{Haynes-11}  with  stellar   masses  above  $10^{8}M_\odot$  and
redshifts  below $0.06$.  As  can be  seen, the  majority of  the grey
points in the right panel lie above the relation.

The  H{\sc   i}  plane  given   by  the  new  estimator   in  equation
(\ref{eqn:hiplane}) is plotted  in red open circles or red dots in both 
panels of Figure~\ref{fig:hiplane}.   There  is  rather  little change  for  the
majority of galaxies  in the GASS sample.  However,  the H{\sc i}-rich
galaxies  in the  ALFALFA  sample that  were  previously displaced  to
higher-than-predicted H{\sc i} mass  fractions, are now mostly located
well within the $1\sigma$ region of the new relation.

We note that this reduction in the systematic offset for H{\sc i}-rich
galaxies could not  be achieved by introducing a  single new parameter
into the fit  (i.e only $\Delta_{g-i}$).  Equation (\ref{eqn:hiplane})
implies  that the  predicted gas  fraction scales  more  strongly with
colour gradient in high mass  galaxies than in low mass galaxies.  The
most  likely reason  for this  is  that massive  galaxies have  larger
bulge-to-disk  ratios than  less massive  galaxies. \citet{Fabello-11}
showed that  the H{\sc i}  content of a  galaxy did not depend  on its
bulge-to-disk ratio; the  H{\sc i} mass fraction only  depended on the
properties of the disk.  It is  thus likely that the H{\sc i} fraction
correlates with the colour gradient of the {\em disk} and the bulge is
a contaminant when determining the  colour gradient. At present, we do
not have bulge/disk decompositions for all the galaxies in our sample,
so  we do  not investigate  this  hypothesis in  more depth.   Another
effect that  may be  important is that  massive galaxies  contain more
dust, and  this may  change the relation  between colour  gradient and
H{\sc i} fraction.

We  now  test  whether   our  new  estimator  exhibits  any  remaining
systematic  biases  by  checking   whether  the  {\em  residuals}  are
correlated    with     any    intrinsic    galaxy     property.     In
Figure~\ref{fig:residuals} we plot the residuals for the C10 estimator
(grey dots)  and for the  new estimator  (colourful dots) as  a function  of $M_\ast$,
$\mu_\ast$, $NUV-r$, and $R_{90}/R_{50}$, We only show results for the
ALFALFA sample, where the new  estimator does change the H{\sc i} mass
fraction predictions by a significant amount.

Figure~\ref{fig:residuals}  shows that  the new  estimator leads  to a
significant  reduction in the  large positive  residuals for  H{\sc i}
rich galaxies with low masses and stellar surface mass densities, blue
colours  and low concentration  indices.  The  new estimator  does not
reduce the residuals for galaxies with high stellar surface densities,
red  colours  and  high  concentration  indices.   There  is  still  a
sub-population  of such  galaxies  that are  H{\sc  i}-rich and  where
equation  (\ref{eqn:hiplane}) fails  to predict  the H{\sc  i} content
accurately.  An example  of  such  a system  is  discussed briefly  in
C10.  In addition, one  might worry  that the  H{\sc i}  mass fraction
estimation may be biased in this regime, because many red galaxies are
not detected in both the GASS and ALFALFA surveys

In the left-hand panel of Figure~\ref{fig:hiplane} we plot the results
of the H{\sc i}  stacking analysis by \citet{Fabello-11}.  By stacking
samples  of a few  hundred galaxies,  \citet{Fabello-11} were  able to
estimate  mean  H{\sc i}  mass  fractions  for  galaxies with  $NUV-r$
colours in the range $4-6$ (shown as green stars on the plot).  As can
be seen,  the C10 estimator accurately reproduces  the stacked results
with  a 1$\sigma$ scatter  less than  0.07 dex,  even for  the reddest
stacks.  Unfortunately,  a similar  test is not  possible for  our new
estimator, because the sample  of SDSS galaxies with available ALFALFA
coverage  is too small  to carry  out a  stacking analysis  using four
different  galaxy  parameters  instead  of two.   Therefore,  in  what
follows,  we divide  our galaxies  into  ``red'' or  ``blue'' using  a
mass-dependent colour divider:
\begin{equation}
 (NUV-r)_{cut} = 0.5\log_{10}(M_\ast/M_\odot)-1.
\end{equation}
 For  ``red''  galaxies  with  $NUV-r>(NUV-r)_{cut}$ we  use  the  old
 estimator, while  the new estimator  is applied to  ``blue'' galaxies
 with $NUV-r<(NUV-r)_{cut}$.

\begin{figure}
  \begin{center}
    \epsfig{figure=f5.ps,clip=true,width=0.48\textwidth}
  \end{center}
  \caption{Logarithm of the mean H{\sc  i} mass fraction is plotted as
    a function  of the logarithm of  the stellar mass  for galaxies in
    the  GASS sample  (black) and  for galaxies  in the  models of
    \citet[][red]{Fu-10}  and \citet[][blue]{Guo-11b}.   The 1$\sigma$
    scatter  in $\log_{10}(M_{HI}/M_\ast)$  is indicated  by  the grey
    shaded area  for the data, and  by red/blue dashed  curves for the
    models.}
  \label{fig:fmh1_vs_mstar}
\end{figure}

\begin{figure*}
  \begin{center}
   \epsfig{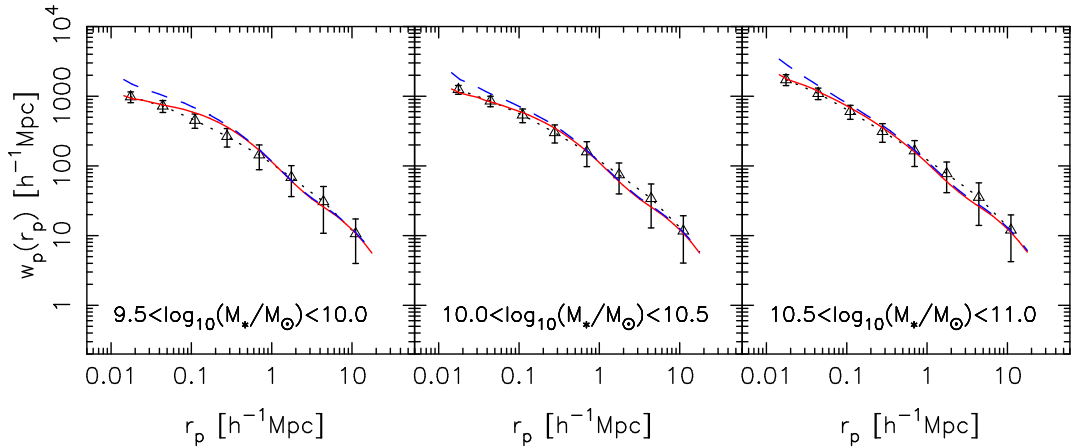}
  \end{center}
\caption{The projected  two-point cross-correlation function  $w_p(r_p)$ 
  for galaxies in bins  of stellar mass with respect to a reference
  sample of all galaxies.  Results for  the SDSS DR7 sample
  are shown in black triangles,  while those for the \citet{Fu-10} and
  \citet{Guo-11b}  models are  shown in  red solid  lines and  in blue
  dashed lines, respectively. Errors on the SDSS results are estimated
  from a  set of 50 mock  galaxy surveys that have  the same selection
  effects as the real SDSS sample.}
\label{fig:wrp_mstar}
\end{figure*}

We now  carry out a test  to see whether the  estimator introduces any
systematic bias in galaxy clustering analyses.  We divide all the GASS
galaxies including  non-detections into six subsamples  of equal size,
using  both the measured  value of  $M_{HI}/M_\ast$ and  the predicted
value  and  we   compute  the  projected  two-point  cross-correlation
functions  (2PCCF),  $w_p(r_p)$  of  these subsamples  with  the  SDSS
reference  sample.  We find  that it  is very  important to  take into
account  the  effect   of  the  errors  in  the   predicted  value  of
$M_{HI}/M_\ast$  when  comparing  the  results using  the  photometric
estimator with the results using the real HI measurements.  The effect
of  errors  is  to weaken  the  clustering  trends  as a  function  of
$M_{HI}/M_\ast$, particularly in the tails of the distribution.  Here,
we model the effect of the  errors by adding a random component to the
measured value of $M_{HI}/M_\ast$ that follows a Gaussian distribution
function with a  width of 0.31 dex.  Results  using the measured H{\sc
  i} mass fractions convolved with the Gaussian distribution of errors
are shown in blue in Figure~\ref{fig:gasswrp}, while results using the
photometric estimator are shown in  red.  The errors in the $w_p(r_p)$
measurements  are computed using  the bootstrap  resampling technique.
The 2PCCF for  the whole GASS sample is plotted as  a black solid line
in each panel for  reference.  Figure~\ref{fig:gasswrp} shows that the
two $w_p(r_p)$ calculations agree well with each other. We have repeated
the same analysis, using the measured H{\sc i} mass fraction for the GASS
galaxies without including the effect of the errors, and found the 
results change very little, indicating that the smearing by the H{\sc i}
predictor on the correlation function is sufficiently small.

Although the  GASS sample  is small,  we can still  see that  both the
amplitude and  shape of the  2PCCF show strong systematic  trends with
increasing H{\sc  i} mass fraction.   H{\sc i}-rich galaxies  are less
strongly  clustered on  all  scales, with  more  pronounced 1-halo  to
2-halo transitions at $\sim 1$Mpc.  Since galaxy clustering depends on
a variety of physical properties, in particular on stellar mass, it is
unclear to  what extent the effect  seen from Figure~\ref{fig:gasswrp}
is due  to H{\sc i}  content only. We  will address this point  in the
next section.

\section{Clustering as a function of M$_{HI}$/M$_\ast$ and comparisons 
with semi-analytic models}

In this  section we  apply our new  photometric estimator to  our full
SDSS DR7 galaxy sample to  study the dependence of clustering on H{\sc
  i} mass fraction.  We compare our results with  predictions from the
G11 and F10 models.  It is well known that clustering depends strongly
on  galaxy stellar mass,  so the  analyses are  always carried  out in
narrow  mass intervals.  In order  to take  errors in  the photometric
estimator  into  account, we  convolve  the  H{\sc  i} mass  fractions
predicted by the models with a Gaussian distribution function of width
0.3 dex in $\log_{10}(M_{HI}/M_\ast)$.

Before  we begin,  we demonstrate  that the  models  reproduce average
trends  in  H{\sc i}  mass  fraction as  a  function  of stellar  mass
reasonably well.   In Figure \ref{fig:fmh1_vs_mstar},  the black solid
curve shows the median value of $\log_{10}M_{HI}/M_\ast$ as a function
of stellar mass for galaxies in the GASS survey, while the grey shaded
region  indicates  the 16$^{th}$  to  84$^{th}$  percentile ranges  of
$\log_{10}M_{HI}/M_\ast$.   Note that  the galaxies  without  H{\sc i}
line  detections are  assigned an  H{\sc i}  mass equal  to  the upper
limit.  This is why the black  curve and the shaded region do not fall
below $\log_{10}M_{HI}/M_\ast \sim -1.82$  (see C10 for details on the
detection limits of the survey).  We now perform the same analysis for
the simulated galaxies  and the results are shown in  red and blue for
the  F10 and  G11 models,  respectively.  We  see that  the  G11 model
yields  a higher  median  value of  H{\sc  i} gas  mass
fraction at a given value of  $M_\ast$, when compared to both the data
and the F10 model.

There are two reasons for this:  1) The G11 model does not account for
the partition of  the neutral gas into different  components.  The F10
model includes simple prescriptions for the formation of molecular gas
and also properly  takes into account the contribution  of helium when
making predictions for  H{\sc i} content. 2) The  F10 model parameters
are explicitly  adjusted so  as to match  the H{\sc i}  mass functions
determined  by existing  H{\sc  i} surveys  like  HIPASS and  ALFALFA.
\citet{Kauffmann-12} have shown that  the F10 model can also reproduce
the  distribution of  $M_{HI}/M_\ast$ for  the population  of galaxies
with detectable gas,  but the model does not  provide a fully accurate
description of the population of galaxies without detectable cold gas.
We note that Figure \ref{fig:fmh1_vs_mstar} includes both populations,
so the fit to the data is not as good as that shown in Figure 2 of the
Kauffmann et al. paper.  Since the GASS is a survey only for high mass
galaxies  with  $\log_{10}(M_\ast/M_\odot)>10$,  we  are not  able  to
extend this comparison to lower masses. We have compared the models to
the  data using our  {\em pseudo}  H{\sc i}  mass fractions,  and this
shows that the models roughly match the data at lower masses as well.

Next, Figure  \ref{fig:wrp_mstar} demonstrates  that both the  G11 and
F10 models reproduce the  observed dependence of $w_p(r_p)$ on stellar
mass.  The agreement with observations is equally good for both models
on large scales.   The F10 model appears to  provide a somewhat better
fit to the clustering amplitude on scales below $\sim 1$ Mpc.

\begin{figure*}
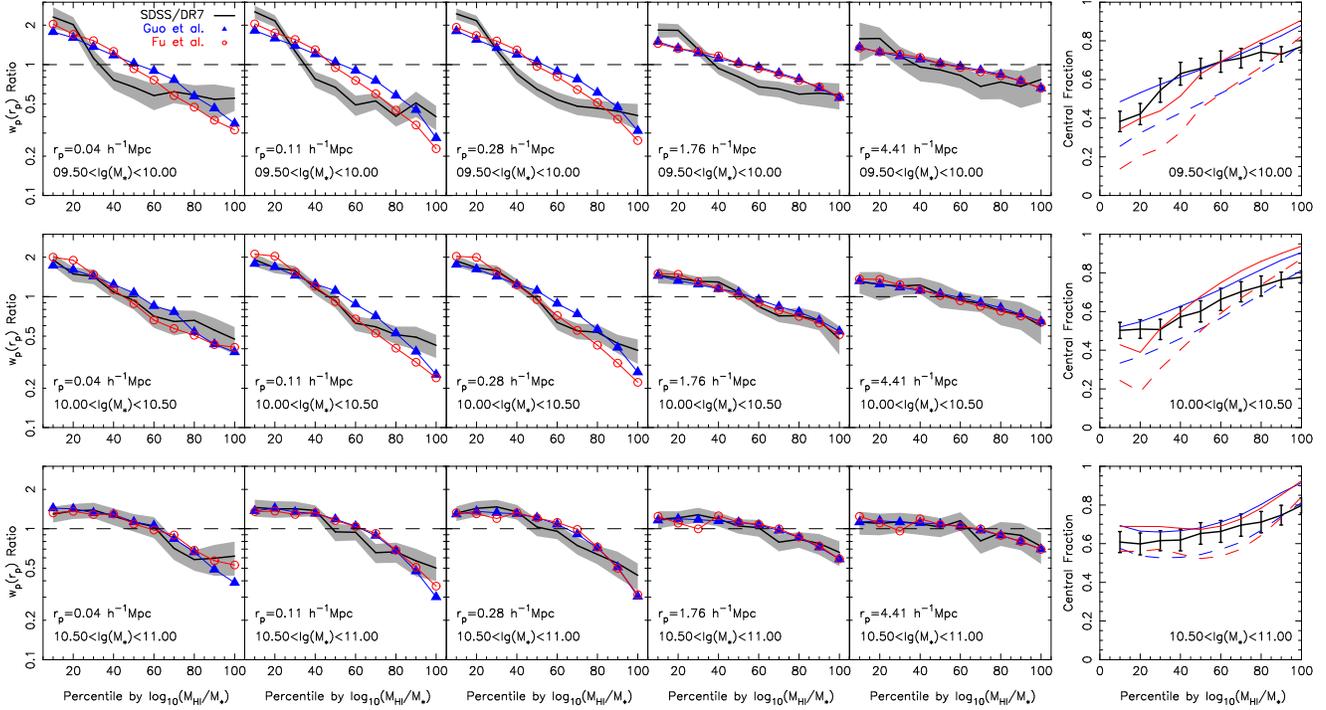

  \begin{center}
    \epsfig{figure=f7a.ps,clip=true,height=0.53\textwidth}
    \epsfig{figure=f7b.ps,clip=true,height=0.53\textwidth}
  \end{center}
  \caption{  Panels on  the left-hand  side show  the  clustering bias
    factor as a function of percentile in $\log_{10}M_{HI}/M_\ast$, at
    different  separations  (panels  from   left  to  right)  and  for
    different stellar  mass ranges (panels  from top to  bottom).  The
    solid line shows observed bias factors, while the shaded region 
    indicate the  errors on the
    observed bias  factors estimated  from the mock  catalogues. Model
    results  are shown  in red/blue  for the  F10/G11 models.   In the
    right-most  column,  the  central  galaxy  fraction is plotted as  a
    function  of  percentile  in $\log_{10}M_{HI}/M_\ast$.  The  black
    curves with error bars show the central  fractions estimated  from 
    the  data (see text). The dashed red/blue  curves show the true central 
    fractions from  the  models. The  solid  red/blue  curves  show the  
    central fractions in the  models when estimated in the same  way as 
    in the data.}
\label{fig:sdsswrp_new}
\end{figure*}

In spite of the good agreement as shown above, there still exists 
discrepancies in some cases. 
In order to carry out  meaningful comparisons between data and models,
we order all the galaxies in  a given stellar mass range by increasing
$M_{HI}/M_\ast$  and  divide the  galaxies  into  10 subsamples,  each
containing 10 per cent of  the whole sample. We analyze the dependence
of $w_p(r_p)$ on H{\sc i} mass fraction as a function of {\em H{\sc i}
  mass fraction percentile} instead of  the absolute value of H{\sc i}
mass fractions\footnote{We note that this only makes sense for the G11
  model if  the intrinsic scatter  in H{\sc i}-to-H$_2$ ratio  in real
  galaxies does not change the ranking of $M_{HI}/M_\ast$ with respect
  to  $[M_{HI}+M_{H_2}]/M_\ast$.  \citet{Saintonge-11b} show  that the
  average value  of $M_{HI}/M_{H_2}$ is  around 1/3 and  the molecular
  gas mass very rarely exceeds the  atomic gas mass, so this is likely
  to be close to correct.}  In  order to provide a more intuitive feel
for our  results, we  present our measurements  in terms of  {\em bias
  factor}, defined  as the ratio of  the $w_p(r_p)$ for  a given H{\sc
  i}-selected  subsample to  the  $w_p(r_p)$ of  all  galaxies in  the
corresponding stellar mass range.  In Figure~\ref{fig:sdsswrp_new}, we
plot   this   bias   factor   as   a   function   of   percentile   in
$\log_{10}M_{HI}/M_\ast$, with H{\sc  i} mass fraction increasing from
left to  right.  Results for  different intervals in stellar  mass are
shown  in  different  rows,   while  results  evaluated  on  different
projected scales  $r_p$ are shown  in different columns.  The  data is
shown  in black curves  with shaded  regions indicating  the 1$\sigma$
errors that are estimated from  the bias factor measurements of the 50
mock SDSS  catalogues, while the F10  and G11 models are  shown in red
circles and blue triangles, respectively.

As can  be seen,  the bias always  decreases with increasing  H{\sc i}
mass fraction.   The trend is strongest  for low mass  galaxies and on
scales  of  around  100-200   h$^{-1}$  kpc.   The  agreement  between
observations  and   models  is  good,  except  for   the  lowest  mass
galaxies. For galaxies with stellar masses in the range $3 \times 10^9
M_{\odot}$-  $10^{10}  M_{\odot}$,  the  observed  bias  factor  drops
steeply as a  function of H{\sc i} mass  fraction percentile, and then
flattens. The  bias factor  predicted by the  models exhibits  a more
linear trend as a function of H{\sc i} mass fraction percentile.
  
As we  will now  show, this decrease  in bias  can be understood  in a
simple  way in terms  of an  increasing ratio  of central-to-satellite
galaxies as a function of increasing H{\sc i} mass fraction.  To prove
that this is the case, we classify each galaxy in our sample as either
a central  galaxy or a  satellite galaxy based  on whether it  is more
massive than  all companions within  a cylinder with  projected radius
$R_{max}$ and  a line-of-sight depth of $\pm1000$  km s$^{-1}$.  Here,
$R_{max}$ is  set to twice the  virial radius of the  host dark matter
halo  of the  galaxy.   We  have adopted  the  stellar mass-halo  mass
relation derived  by \citet{Guo-10}  to estimate a  halo mass  for the
galaxy,  and then estimate  a `virial'  radius of  the halo  using the
model  of \citet{Eke-Navarro-Steinmetz-01}.   In addition,  we require
that a  central galaxy should not  fall within $R_{max}$  of any other
more massive galaxy.

In the right-hand panels  in Figure~\ref{fig:sdsswrp_new}, we plot the
fraction of  central galaxies,  $f_{cen}$, as a  function of  H{\sc i}
fraction  percentile (the black  solid line).   We see  that $f_{cen}$
increases with  increasing H{\sc i} content, with  the effect stronger
at low stellar masses.

We can of  course check whether the models  predict a similar increase
in $f_{cen}$  as a function  of H{\sc i}  gas fraction.  In  the right
column  of Figure~\ref{fig:sdsswrp_new},  the dashed  curves  show the
{\em  true} values of  $f_{cen}$ as  a function  of H{\sc  i} fraction
percentile  for the  F10 (red)  and  G11 (blue)  models.  The  central
fractions in  the F10 model are lower  than those in the  G11 model at
low  H{\sc  i} mass  fractions,  particularly  for  galaxies with  low
stellar masses. This  reflects the fact that gas  consumption times in
satellite galaxies are longer in the G11 model than in the F10 model.

In order  to make  a fair comparison  with observations, we  have also
computed $f_{cen}$ for  the model galaxies in exactly  the same way as
in the observations.  In brief, we project the model galaxies onto the
$x-y$ plane and take the $z$-axis as the line-of-sight direction (i.e.
we adopt  the distant observer approximation).  We  then apply exactly
the same procedure described above  to classify each galaxy as central
or satellite.   The $z-$axis peculiar  velocities of the  galaxies are
added  to their  $z$-axis  positions.  In  addition,  halo masses  and
virial radii are not taken from the model catalogue, but are estimated
exactly  the same  way as  for  the observational  data.  Results  are
plotted  as solid  red and  blue curves  in the  right-hand  panels of
Figure~\ref{fig:sdsswrp_new}.

We note  that the true central  fractions are always  smaller than the
ones  that use  a classification  technique  based on  whether or  not
brighter companions are found in cylinders around the galaxy. However,
our  classification  technique preserves  the  shape  of the  relation
between central fraction and H{\sc i} gas fraction percentile, as well
as the differences  between the F10 and G11 models.   We also see that
the central  fractions estimated in cylinders in  the simulation agree
reasonably well  with the data. As  was the case for  the bias factor,
the behaviour  of the  central fraction as  function of H{\sc  i} mass
fraction percentile  in the  models is somewhat  different to  what is
seen in the observations, particularly at low stellar masses.

In  summary,  the  general  agreement  with the  models  supports  our
conjecture that  the trends in bias  factor as a function  of H{\sc i}
mass   fraction  mainly   arise  as   a  result   of  trends   in  the
satellite-to-central ratio.

\section {Scale dependence of the bias for galaxies with 
excess/deficient H{\sc i} content}

\begin{figure*}
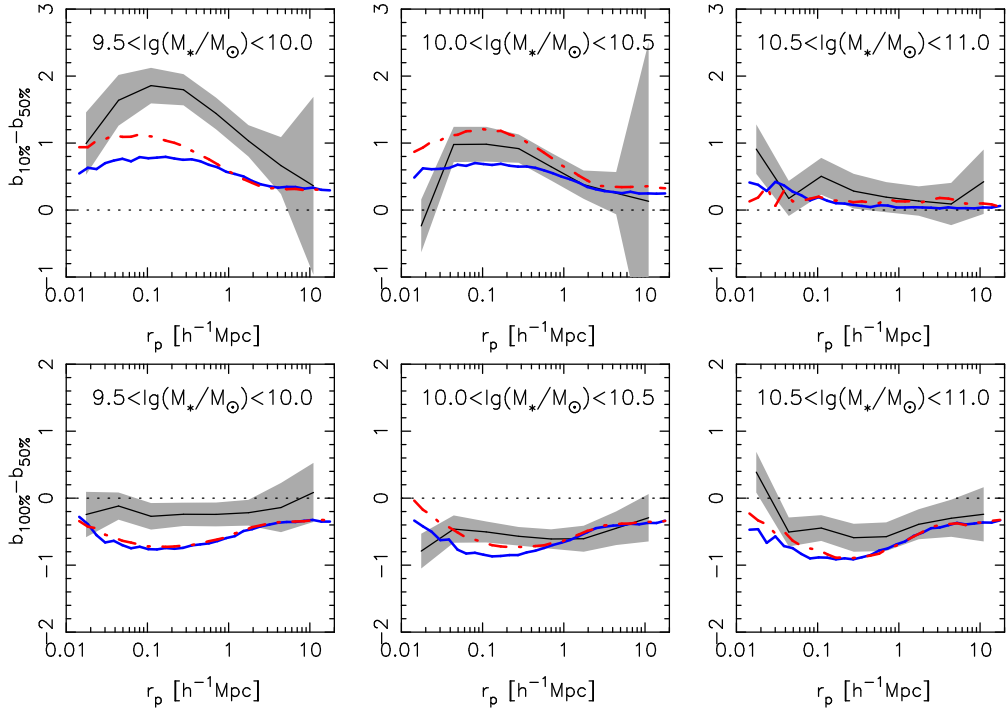

  \begin{center}
    \epsfig{figure=f8a.ps,clip=true,width=0.75\textwidth}
    \epsfig{figure=f8b.ps,clip=true,width=0.75\textwidth}
  \end{center}
  \caption{In the upper panels, we plot the change in bias factor from
    the   10$^{th}$  percentile   to  the   50$^{th}$   percentile  in
    $\log_{10}(M_{HI}/M_\ast)$  as a  function  of projected  physical
    scale,  for different  stellar  mass intervals  as indicated.  The
    black  line  shows  the  result  from  the  SDSS/DR7.  The  errors
    estimated   from  the   mock  catalogues   are  shown   as  shaded
    regions. The red  dotted-dashed line and the blue  solid line show
    results  from the \citet{Fu-10}  and \citet{Guo-11b}  models after
    convolution with errors. The lower  panels show the change in bias
    factor from the 50$^{th}$ to the 100$^{th}$ percentile.}
  \label{fig:scalebias1}
\end{figure*}

\begin{figure*}
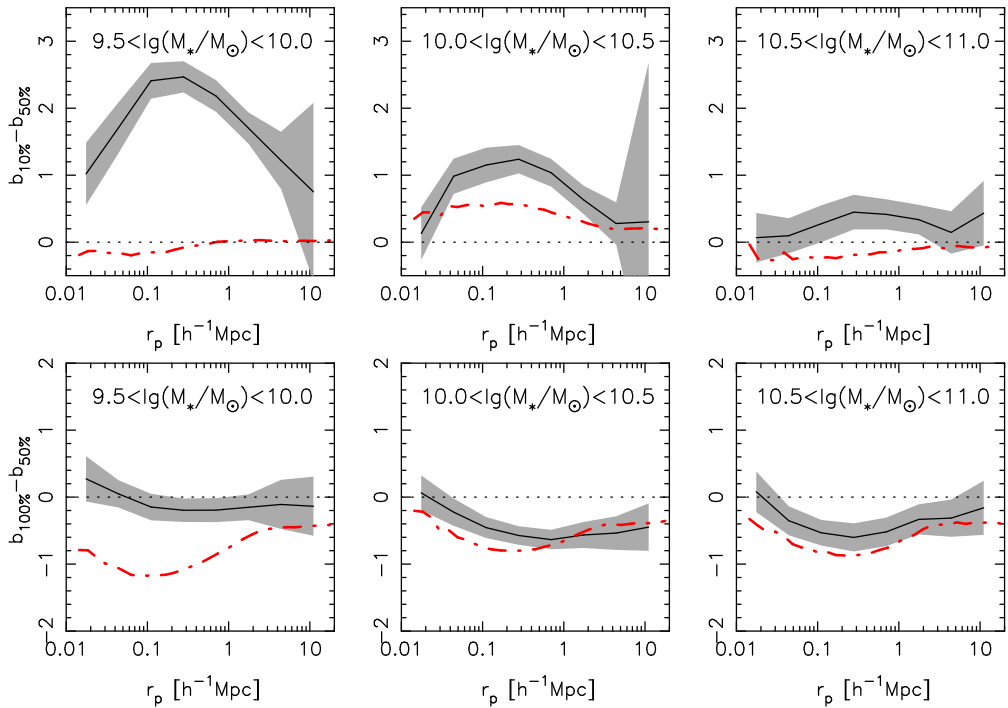

  \begin{center}
    \epsfig{figure=f9a.ps,clip=true,width=0.75\textwidth}
    \epsfig{figure=f9b.ps,clip=true,width=0.75\textwidth}
  \end{center}
  \caption{Same as  the previous figure, except that  the galaxies are
    ordered  by the deviation  in $\log_{10}(M_{HI}/M_\ast)$  from the
    value    predicted     from    the    mean     relation    between
    $\log_{10}(M_{HI}/M_\ast)$  and galaxy  mass $M_\ast$  and stellar
    surface  mass density  $\mu_\ast$. Red  dotted-dashed  curves show
    results  from the  model of  \citet{Fu-10} after  convolution with
    errors.}
  \label{fig:scalebias2}
\end{figure*}

\begin{figure*}
  \begin{center}
    \epsfig{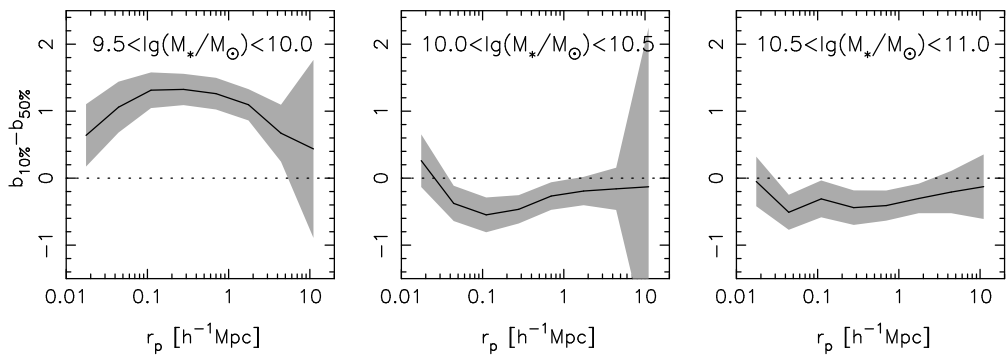}
  \end{center}
  \caption{Same  as the bottom  panel of  the previous  figure, except
    that   the   galaxies   are    ordered   by   the   deviation   in
    $\log_{10}(M_{HI}/M_\ast)$ from the  value predicted from the mean
    relation   between  $\log_{10}(M_{HI}/M_\ast)$  and   galaxy  mass
    $M_\ast$,  stellar surface  mass density  $\mu_\ast$,  and $NUV-r$
    colour.}
  \label{fig:scalebias3}
\end{figure*}

In the previous  section, we studied how the bias  factor changes as a
function  of  normalized  H{\sc  i}  mass  fraction  for  galaxies  in
different  stellar  mass  bins.    The  results  presented  in  Figure
~\ref{fig:sdsswrp_new} clearly show that the change in the bias factor
between ``gas-poor'' and ``gas-rich'' galaxies depends on scale $r_p$.

In this section,  we analyze the scale-dependence of  the {\em change}
in bias factor for both  gas-deficient and gas-rich galaxies.  We note
that  \citet{Haynes-Giovanelli-84} defined  gas-deficiency  to be  the
difference in H{\sc i}  content between cluster galaxies and ``field''
galaxies  of the  same morphological  type and  size.  There  are some
difficulties  with  this   definition,  including  the  definition  of
``field'' and  the fact that  Hubble classification is  problematic in
rich clusters. Some more  recent analyses have used a type-independent
deficiency parameter  that compares all  galaxies to a fixed  H{\sc i}
surface  density \citep{Chung-09}. One  worry with  this, is  that the
mean  H{\sc  i}  content  of  galaxies  scales  strongly  with  galaxy
parameters    such    as   stellar    mass    and   surface    density
\citep{Catinella-10}.

In  this analysis,  we  will  adopt a  flexible  approach to  defining
``pseudo''  H{\sc  i}  deficiency  parameters.  We  will  analyze  the
photometrically-predicted H{\sc i} content with respect to galaxies of
the same stellar  mass, with respect to galaxies of  the same mass and
size, and with  respect to galaxies of the same  mass, size and colour
(as in \citealt{Cortese-11}).

In  the upper  panels  of Figure  ~\ref{fig:scalebias1},  we show  the
change in  bias factor between the 10$^{th}$  and 50$^{th}$ percentile
bins  in   $\log_{10}M_{HI}/M_\ast$.   This   serves  as  a   test  of
gas-stripping mechanisms in  gas-deficient galaxies.  The lower panels
show the  change in bias  factor between the 100$^{th}$  and 50$^{th}$
percentile bins in $\log_{10}M_{HI}/M_\ast$.  This serves as a test of
gas accretion mechanisms in gas-rich galaxies.

Results for the SDSS DR7 galaxies  are shown as black curves in Figure
~\ref{fig:scalebias1}.   Grey shaded  regions  indicate the  1$\sigma$
errors   on  our   estimates,   obtained  from   the   50  mock   SDSS
catalogues. Results  for the F10 and  G11 models are shown  as red and
blue curves  and we plot our  results in three  different stellar mass
ranges.   As can  be seen  from the  plot, the  change in  bias factor
between gas-deficient galaxies and galaxies with typical gas fractions
is  most pronounced  for low  stellar  mass systems.  The bias  factor
difference peaks  at relatively small physical  scales ($\sim 100-300$
kpc).      For      the      most      massive      galaxies      with
$10.5<\log_{10}(M_\ast/M_\odot)<11$ ,  there is little  change in bias
on any  scale.  For galaxies  with $9.5<\log_{10}(M_\ast/M_\odot)<10$,
the increase  in clustering amplitude  from galaxies with  typical gas
mass fractions to the most gas-deficient objects reaches a factor of 2
on scales  of a  few hundred  kpc.  On scales  larger than  $2-3$ Mpc,
there is  no significant change in clustering  amplitude.  The results
are consistent with the idea that gas quenching is driven by processes
that are {\em  internal} to dark matter halos.   The models agree well
with the data at stellar  masses greater than $10^{10} M_{\odot}$, but
at  lower  stellar  masses  the  models  predict  a  weaker  bias  for
gas-deficient galaxies than is actually seen.

As  seen in  the bottom  panels of  Figure  ~\ref{fig:scalebias1}, the
change in bias factor between very gas-rich galaxies and galaxies with
typical gas fractions appears to be weaker rather than stronger at low
stellar  masses.   The  most  gas-rich galaxies  with  stellar  masses
greater  than  $10^{10}  M_{\odot}$  are more  weakly  clustered  than
galaxies with typical gas fractions, indicating that they occupy lower
mass dark  matter halos on  average. At stellar masses  below $10^{10}
M_{\odot}$, there  is no  anti-bias of gas-rich  galaxies seen  in the
data. However, the models do predict clear anti-bias effects.

One might question whether ranking  galaxies by H{\sc i} mass fraction 
is sufficient  to characterize  whether a  galaxy is  classified as
gas-rich or gas-deficient.   As discussed in \S~\ref{sec:hiestimator},
galaxies of  fixed stellar mass and  colour have higher  H{\sc i} mass
fractions if they  have larger sizes (i.e. lower  stellar surface mass
densities). One  way to  understand this  is to appeal to standard  disk
formation models \citep[e.g. F10;][]{Kauffmann-96a, Mo-Mao-White-98}.
In these models,  the  spin parameter of  the dark  matter halo
determines   the    contraction   factor   of    the   infalling   gas.
Larger  disks in a dark matter halo of fixed mass   will  have
higher H{\sc i}  mass fractions because gas surface  densities are low
and  gas  consumption times  are  long. In this case, it would make
more sense to define galaxies as gas-rich or gas-deficient by comparing
their H{\sc i} mass fractions to other galaxies of the same {\em mass and size}.    

One might also  consider an even more stringent  constraint that H{\sc
  i}-rich/H{\sc i}-deficient  galaxies be classified  as those objects
with  higher/lower-than-average H{\sc i}  content given  their stellar
mass,  size and  star formation  rate.  This might  indicate that  the
galaxy has  experienced a  recent gas accretion  episode and  that the
{\em global} star formation has not yet had a chance to respond to the
extra fuel supply.   In our scheme of using  photometric quantities to
predict H{\sc  i} content, the H{\sc i}-rich  systems would correspond
to those galaxies with bluer-than-average outer disks. Recall that the
H{\sc i} content in gas-poor regime is currently calibrated using only
stellar surface density and colour;  we therefor do not delve into the
opposite regime, where gas has been recently removed from a galaxy.
  
In   Figures   ~\ref{fig:scalebias2}   and  ~\ref{fig:scalebias3}   we
investigate clustering trends using these alternative definitions. For
Figure ~\ref{fig:scalebias2}, we rank  galaxies as a function of their
deviation from the  average H{\sc i} mass fraction  of all galaxies of
the  same stellar  mass ($M_\ast$)  and stellar  surface  mass density
($\mu_\ast$).    As  seen   from   equation~(\ref{eqn:hiplane}),  this
deviation depends  on both  the $NUV-r$ colour  of the galaxy  and its
$g-i$  colour  gradient.  For  Figure  ~\ref{fig:scalebias3}, we  rank
galaxies as  a function of their  deviation from the  average H{\sc i}
mass fraction  of all  galaxies of the  same $M_\ast$,  $\mu_\ast$ and
$NUV-r$.  This then  depends only on the $g-i$  colour gradient of the
galaxy.

Interestingly,  the top  panels of  Figure  ~\ref{fig:scalebias2} show
that when gas deficiency is expressed relative to galaxies of the same
stellar mass  and size,  the change  in bias on  scales between  a few
hundred kpc  and 1  Mpc becomes much  more pronounced.  The  change in
bias factor for the lowest mass galaxies now reaches values near $\sim
3$  and  even massive  gas-deficient  galaxies  are now  significantly
biased  with  respect  to   their  counterparts  with  ``normal''  gas
fractions.

The F10 model  provides predictions of the radial  profiles of the gas
and  the stars  in  galaxies.  It  is  thus possible  to  look at  gas
deficiency at  fixed mass  and stellar surface  density in  the model.
Results are  plotted as red curves  in Figure~\ref{fig:scalebias2}. We
see that the  model disagrees very strongly with  the observations. In
the model, bias effects become  {\em weaker rather than stronger} when
gas deficiency  is defined with respect  to galaxies of  the same mass
and  size. These  results would  appear to  suggest that  in  the real
Universe,  gas removal  processes depend  on the  size/density  of the
galaxy itself.   This is not the  case in the  models, where satellite
galaxies become  gas-poor only because  their supply of  infalling gas
has  been cut  off.   Thus the  data  suggest that  processes such  as
ram-pressure   stripping,  which   depend  on   the  density   of  the
interstellar medium  (ISM), may play  an important part  in explaining
the observed trends.
  
In  contrast   to  what  is   seen  for  gas-deficient   galaxies,  if
gas-richness  is  normalized with  respect  to  galaxies  of the  same
stellar mass  and size, the bias  trends remain much the  same and the
F10  model predictions  still fit  reasonably well  for  galaxies more
massive  than  $10^{10}M_\odot$.  This  suggests  that  gas  accretion
processes  are  being  modelled  quite successfully  at  high  stellar
masses.

Figure ~\ref{fig:scalebias3} shows that  the bias effects for gas-rich
galaxies are  still roughly  of the same  strength as in  the previous
figure,  when gas-richness is  expressed relative  to galaxies  of the
same  mass,  size and  global  $NUV-r$  colour.  This means  that  the
clustering does depend strongly  on the colour-gradient term for these
objects.  Galaxies  with bluer-than-average outer  colours are clearly
located in lower-density environments compared to galaxies where there
is  no evidence  for younger-than-average  stellar populations  in the
outer disk.

\section{Summary and discussion}

We introduce a  new photometric estimator for estimating  the H{\sc i}
mass fraction  ($M_{HI}/M_\ast$) in local galaxies.   The estimator is
calibrated with a sample  of 293 galaxies with $M_\ast>10^{10}M_\odot$
in the redshift  range $0.025<z<0.05$, which are detected  in H{\sc i}
emission  line  by  the  GASS   survey.  The  estimator  is  a  linear
combination of four parameters: stellar mass $M_\ast$, stellar surface
mass  density $\mu_\ast$, near-UV-to-optical  colour $NUV-r$,  and the
gradient  in $g-i$  colour $\Delta_{g-i}$.   We demonstrate  that this
estimator  provides  unbiased   $M_{HI}/M_\ast$  estimates  for  H{\sc
i}-rich galaxies.

We then apply this estimator to a sample of $\sim$24,000 galaxies from
the  SDSS/DR7 that  lie in  the same  redshift range.  We  analyze the
clustering of  these galaxies as a  function of stellar mass  and as a
function of H{\sc i} mass  fraction $M_{HI}/M_\ast$ and we compare the
results  with  predictions from  two  recent  semi-analytic models  of
galaxy formation by \citet{Fu-10} and \citet{Guo-11b}. Our results may
be summarized as follows:

\begin {itemize}
\item
Clustering  depends  strongly  on  H{\sc  i} mass  fraction  at  fixed
$M_\ast$.   Galaxies with  large  values of  $M_{HI}/M_\ast$ are  more
weakly clustered.   The total  change with H{\sc  i} mass  fraction in
clustering strength is largest for low mass galaxies,
\item
At fixed $M_\ast$, the clustering dependence on H{\sc i} mass fraction
is strongest  on scales of  a few hundred  kpc.  On large  scales ($>$
1Mpc),  clustering depends  weakly on  H{\sc i}  mass  fraction.  This
suggests  the H{\sc  i}  content of  a  galaxy of  fixed stellar  mass
depends on location within its dark matter halo.
\item
After the uncertainty in the H{\sc i} mass fraction estimator is taken
into account, the  observed dependence of clustering on  H{\sc i} mass
fraction is  well reproduced by  the models for galaxies  more massive
than  $10^{10}M_\odot$.  Significant  discrepancies  remain  at  lower
stellar masses.
\end {itemize}

In the next part of the  paper, we extend the analysis by studying the
clustering of H{\sc i}-deficient and H{\sc i}-rich galaxies defined in
two ways: 1) with respect to  the average H{\sc i} content of galaxies
of the  same stellar  mass and  size, 2) with  respect to  the average
H{\sc  i} content  of  galaxies of  the  same stellar  mass, size  and
$NUV-r$  colour.  These  definitions  are motivated  by the  following
considerations.  First, models in which disks form from gas that cools
and condenses in dark matter halos, while conserving angular momentum,
predict that  the gas  fractions of equilibrium  disks depend  on both
their  mass and  their size.   Second, the  majority of  nearby spiral
galaxies are {\em observed} to lie on a relatively tight plane linking
H{\sc i} gas mass fraction  with stellar mass, stellar surface density
and $NUV-r$ colour \citep[][]{Catinella-10}.  It is natural to suppose
that galaxies that  have undergone a recent gas  accretion event would
be displaced  to higher H{\sc i}  mass fractions with  respect to this
plane. Conversely, galaxies that have been stripped of their gas would
be displaced to lower values of $M_{HI}/M_\ast$.

The  main  results  of  our   analysis  of  H{\sc  i}-rich  and  H{\sc
i}-deficient galaxies can be summarized as follows.

\begin {itemize}
\item
When H{\sc  i} deficiency is defined  with respect to  galaxies of the
same stellar  mass and size,  the bias of H{\sc  i}-deficient galaxies
relative to  normal galaxies is larger  than obtained if  the H{\sc i}
deficiency is defined  with respect to galaxies of  the same mass. The
same  effect   is  not  reproduced  in  the   semi-analytic  model  of
\citet{Fu-10}.
\item
When H{\sc  i}-richness is expressed  with respect to galaxies  of the
same mass  and size (as well as  with respect to galaxies  of the same
mass,  size and  colour),  H{\sc i}-rich  galaxies  more massive  than
$10^{10}M_\odot$ are observed to  be anti-biased with respect to their
counterparts with  normal H{\sc  i} content. The  same is not  true at
lower stellar masses.
\end {itemize}

We have  proposed that the  disagreement between the  observations and
the  models  might be  resolved,  if  processes  such as  ram-pressure
stripping, which depend on the density of the ISM, are included in the
models.  We  note  that  the  lowest mass  galaxies  have  the  lowest
densities and are thus the  most likely to be affected by ram-pressure.
In  order to  test this  hypothesis in  more detail,  we plan  to look
behaviour of gas deficiency as a function of cluster-centric radius in
samples of nearby groups and clusters.

We also stress  that next generation wide-field H{\sc  i} surveys such
as the ASKAP H{\sc i} All-sky Survey (WALLABY) and surveys carried out
by  the  Apertif receiver  array  on  the  Westerbork Synthesis  Radio
Telescope will measure  H{\sc i} masses and sizes  for samples of tens
to  hundreds of  thousands of  galaxies  at redshifts  of around  0.1.
These surveys  will make  it possible to  investigate clustering  as a
function  of  the true  H{\sc  i}  content of  a  galaxy.  It will  be
interesting  to investigate the  degree to  which the  results conform
with   our  current   analysis   of  ``pseudo''   H{\sc  i}   content.
Discrepancies  may reveal  additional  physics that  we  have not  yet
considered.   In the  meantime, the  construction of  models  that can
reproduce the  gas properties of galaxies  as well as  possible, is an
important step towards  building mock surveys that can  be safely used
for making predictions in support of these surveys.

\section*{Acknowledgments}
CL  thanks the Max-Planck  Institute for  Astrophysics (MPA)  for warm
hospitality  while  this  work  was  being completed.   This  work  is
sponsored   by  NSFC   (no.   11173045),   Shanghai   Pujiang  Program
(no. 11PJ1411600) and  the CAS/SAFEA International Partnership Program
for  Creative  Research  Teams  (KJCX2-YW-T23).  CL  acknowledges  the
support  of the  100 Talents  Program of  Chinese Academy  of Sciences
(CAS) and the exchange program between Max Planck Society and CAS.

Funding for  the SDSS and SDSS-II  has been provided by  the Alfred P.
Sloan Foundation, the Participating Institutions, the National Science
Foundation, the  U.S.  Department of Energy,  the National Aeronautics
and Space Administration, the  Japanese Monbukagakusho, the Max Planck
Society,  and the Higher  Education Funding  Council for  England. The
SDSS Web  Site is  http://www.sdss.org/.  The SDSS  is managed  by the
Astrophysical    Research    Consortium    for    the    Participating
Institutions. The  Participating Institutions are  the American Museum
of  Natural History,  Astrophysical Institute  Potsdam,  University of
Basel,  University  of  Cambridge,  Case Western  Reserve  University,
University of Chicago, Drexel  University, Fermilab, the Institute for
Advanced   Study,  the  Japan   Participation  Group,   Johns  Hopkins
University, the  Joint Institute  for Nuclear Astrophysics,  the Kavli
Institute  for   Particle  Astrophysics  and   Cosmology,  the  Korean
Scientist Group, the Chinese  Academy of Sciences (LAMOST), Los Alamos
National  Laboratory, the  Max-Planck-Institute for  Astronomy (MPIA),
the  Max-Planck-Institute  for Astrophysics  (MPA),  New Mexico  State
University,   Ohio  State   University,   University  of   Pittsburgh,
University  of  Portsmouth, Princeton  University,  the United  States
Naval Observatory, and the University of Washington.

\bibliography{ref}

\bsp
\label{lastpage}
\end{document}